\def\year{2022}\relax
\documentclass[letterpaper]{article} 
\usepackage{aaai22}  
\usepackage{times}  
\usepackage{helvet}  
\usepackage{courier}  
\usepackage[hyphens]{url}  
\usepackage{graphicx} 
\usepackage{natbib}  
\usepackage{caption} 
\DeclareCaptionStyle{ruled}{labelfont=normalfont,labelsep=colon,strut=off} 
\usepackage{booktabs}
\usepackage{algorithm}
\usepackage{algorithmic}
\usepackage{multirow}
\usepackage{subfigure}

\usepackage{amssymb}
\usepackage{amsfonts}
\usepackage{mathrsfs}
\usepackage{amsmath}
\usepackage{booktabs}
\usepackage{bbm}
\usepackage{algorithm}
\usepackage{algorithmic}
\usepackage{makecell}
\usepackage{multirow}
\usepackage{subfigure}
\usepackage{amsmath}
\usepackage{amssymb}
\usepackage[switch]{lineno}
\usepackage{color}
\usepackage{newfloat}
\usepackage{listings}

\newtheorem{theorem}{Theorem}
\newtheorem{lemma}{Lemma}
\newtheorem{definition}{Definition}
\newtheorem{remark}{Remark}
\newtheorem{proposition}{Proposition}
\newtheorem{condition}{Condition}

\newenvironment{proproof2}{\noindent{\textbf{Proof of Proposition 2}:}}{\quad \hfill$\Box$\vspace{2ex}}
\newenvironment{proproof3}{\noindent{\textbf{Proof of Proposition 3}:}}{\quad \hfill$\Box$\vspace{2ex}}
\newenvironment{prolemma4}{\noindent{\textbf{Proof of Lemma 4}:}}{\quad \hfill$\Box$\vspace{2ex}}

\newenvironment{proof6}{\noindent{\textbf{Proof of Theorem 6}:}}{\quad \hfill$\Box$\vspace{2ex}}

\newenvironment{proof4}{\noindent{\textbf{Proof of Theorem 4}:}}{\quad \hfill$\Box$\vspace{2ex}}

\usepackage{newfloat}
\usepackage{listings}
\floatstyle{ruled}
\newfloat{listing}{tb}{lst}{}
\floatname{listing}{Listing}
\usepackage{color}
\title{Error-based Knockoffs Inference for Controlled Feature Selection}

\author {
    Xuebin Zhao\textsuperscript{\rm 1},
    Hong Chen\textsuperscript{\rm1 ,}\thanks{Corresponding author.},
          Yingjie Wang\textsuperscript{\rm 2},
      Weifu Li\textsuperscript{\rm 1},
      Tieliang Gong\textsuperscript{\rm 3},
      Yulong Wang\textsuperscript{\rm 2},
    Feng Zheng\textsuperscript{\rm 4}\\
}
\affiliations {
  \textsuperscript{\rm 1}College of Science, Huazhong Agricultural University, Wuhan 430062, China\\
\textsuperscript{\rm 2}College of Informatics, Huazhong Agricultural University, Wuhan 430062, China\\
	 \textsuperscript{\rm 3}School of Computer Science and Technology, Xi'an Jiaotong University, Xi'an 710049, China\\	
	 	 \textsuperscript{\rm 4}Department of Computer Science and Engineering, Southern University of Science and  Technology, China\\
    chenh@mail.hzau.edu.cn;
}

\begin{document}

\maketitle

\begin{abstract}
  Recently, the scheme of model-X knockoffs was proposed as a promising solution to address controlled feature selection under high-dimensional finite-sample settings. However, the procedure of model-X knockoffs
  depends heavily on the coefficient-based feature importance and only concerns the control of false discovery rate (FDR). To further improve its adaptivity and flexibility, in this paper,  we propose an error-based knockoff inference method  by integrating the knockoff features, the error-based feature importance statistics, and the stepdown procedure together. The proposed inference procedure does not require specifying a regression model and can handle feature selection with theoretical guarantees on controlling false discovery proportion (FDP), FDR, or $k$-familywise error rate ($k$-FWER).  Empirical evaluations demonstrate the competitive performance of our approach on both simulated and real data.
\end{abstract}

\section{Introduction}
Data-driven feature selection aims to uncover informative features associated with the response to tailor interpretable statistical inference. Based on regression estimation, various regularized models have been formulated for sparse feature selection   \cite{hastie,fan2001variable,Lin-A.S-2007,chenh-icml-2020,Chenh-tnnls-2020}.
	 Following this line, typical methods include  Lasso \cite{Lasso}, group Lasso \cite{group_lasso,Bach-2008,friedman2010a}, LassoNet \cite{LassoNet}, SpAM \cite{Ravikumar-ICNIPS-2007}, GroupSpAM \cite{Yin-ICML-2012}, and regression models with automatic structure discovery  \cite{Chao-NIPS-2017-additive,frecon2018bilevel,chenh-nips-2020}.
It should be noticed that  the above-mentioned works mainly concern the algorithm's power performance to select true informative features.
However, it is  still largely undeveloped to carry out feature selection while explicitly controlling the number of false discoveries \cite{FWER_First_1987,FDR_first_1995,Lehmann_K-FWER_FDP_2005,ModelXknockoff}.

It is well known that false discovery control is crucial to enable interpretable machine learning in many real-world applications, e.g., genetic analysis where the cost of examining a falsely selected gene may be intolerable.  \citet{FWER_First_1987} proposed a method to control the probability of selecting one or more false discoveries, while it may lead to low power in high dimensional settings.  \citet{FDR_first_1995} formulated an approach to control the expect value of false discovery proportion (FDP), which is called FDR control.  To balance the selection accuracy and the power, some trade-off models  \cite{KFWER_FDP_2004,Lehmann_K-FWER_FDP_2005} are constructed for controlling the probability of selecting $k$ or more false discoveries ($k$-FWER control), or the probability of FDP exceeding a fixed level (FDP control). Besides the above works, there are extensive studies on  feature selection with FWER control \cite{review_2008}, FDR control \cite{FDR_2001,FDR_2002,stochastic_process_FDR_2004,FDR2012,FDR2014}, and FDP control \cite{review2010,FDP_2015}. However, most of them either assume a specific dependent structure between the response and argument (such as linear structure) or rely on $p$-value to evaluate the significance of each feature. The structure assumption may be too restrictive in many applications, where the response  could depend on input features through very complicated forms. In addition, the classical $p$-value calculation procedures  usually depend on the
large-sample asymptotic theory, which may be no longer justified under high-dimensional finite-sample settings \cite{ModelXknockoff,P-value-issues-JMLR-2019}.

\subsection{Knockoff Filter}
Recently, novel knockoff statistics have been constructed in \cite{Knockoff_FDR,ModelXknockoff,LuDeepPINK, Knockoff_KIMI,Knockoff_RANK,knockoff_ipad, knockoff_model-free,Knockoff_group_zoom}
 to evaluate the contribution of each feature to the corresponding response. In particular, theoretical analysis demonstrates that irrelevant features' statistics are independent and symmetrically distributed without making any assumption on the sample size, the number of dimensions, or the dependent structure.  This property is then used to discover informative features with FDR control.
 Beyond identifying informative features,  a new testing procedure (called conditional randomization test) is developed in \cite{ModelXknockoff}, which can estimate the distribution of knockoff statistics and construct the valid $p$-values under finite sample settings by repeatedly training the learning model.

Rapid progress has been made in recent years on understanding the theoretical behavior of knockoff techniques. \citet{ModelXknockoff} proved that the model-X knockoffs (MX-Knockoff) framework enjoys tight FDR control when the covariant distribution is known, and feature importance statistic satisfies some mild conditions  (e.g., shown in Proposition \ref{Model-X W_j distribution}). Moreover, some refined works in  \citet{KnockoffTheory2020,Knockoff_RANK,knockoff_ipad} have demonstrated that the knockoff procedures can also control FDR with asymptotic probability one even the covariant distribution follows some unknown Gaussian graphical model. In addition, the power of knockoff filters is guaranteed for RANK \cite{Knockoff_RANK}, IPAD \cite{knockoff_ipad} and \cite{knockoff_power} under the linear model assumption. Recently, a two-step approach has been proposed in \cite{knockoff_model-free} based on the projection correlation and the model-X knockoff features, which has both sure screening and rank consistency under weak assumptions.

\begin{figure}[!t]
\centering
\includegraphics[scale=0.3]{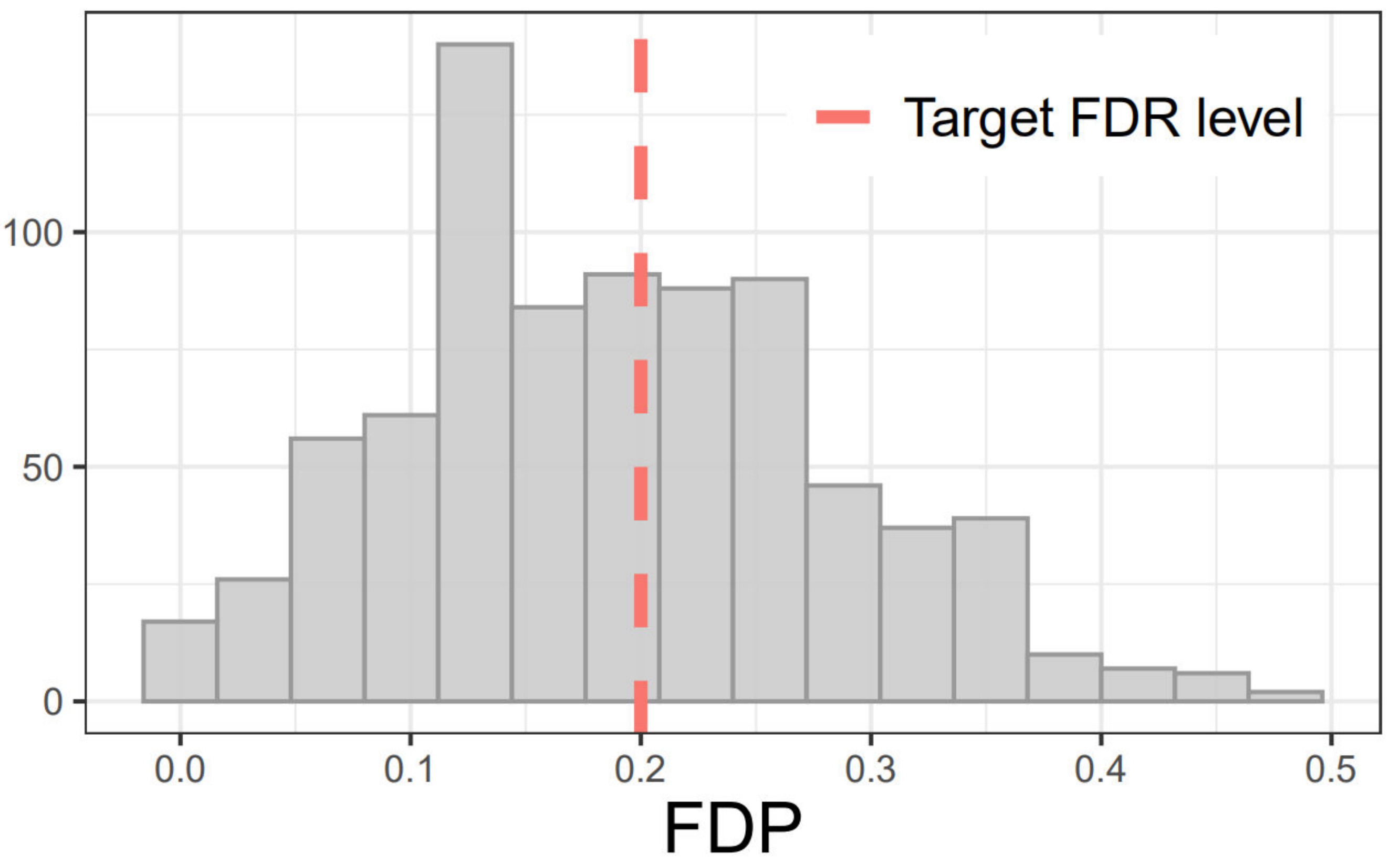}
\caption{FDP under a given target FDR level}
\label{FDPexcueedFDR}
\end{figure}

Despite the success of knockoff filters, some issues remain to be further investigated:
\begin{itemize}
\item \emph{FDR Vs. FDP and $k$-FWER}. {The control of FDR does not assure the control of FDP \cite{stochastic_process_FDR_2004}.}
 To illustrate such phenomenon, we display the histogram of FDP when applying MX-Knockoff \cite{ModelXknockoff} in 800 randomly generated datasets in Figure \ref{FDPexcueedFDR}, which shows FDP can significantly exceed the target FDR level.  See \emph{Supplementary Material B} for details of our simulated example and related discussions.  In addition, as pointed out in \cite{review_2008}, $k$-FWER control is more desirable than FDR control when a powerful selection result can be made.

\item \emph{Coefficient-based feature statistic Vs. Coefficient-free feature statistic}. Under MX-Knockoff framework, feature importance is usually measured by the \emph{coefficient difference}, e.g., \cite{ModelXknockoff,Knockoff_RANK,knockoff_ipad}.
     However, it may be  difficult to obtain the feature importance from general nonlinear models \cite{Ridge,knockoff_model-free}.
    Indeed, it is an open question to design new feature importance statistics (see Section 7.2.5  in \cite{ModelXknockoff}), e.g., coefficient-free statistics.

\item \emph{Conditional randomization test Vs. Computational friendly test}. Although the $p$-value of each feature can be calculated via the conditional randomization test, this procedure is required to train the learning model multiple times. It will cause heavy computational burdens when calculating valid $p$-values, especially in high-dimensional finite-sample case \cite{ModelXknockoff}. Thus, it is an open question  how to efficiently calculate valid $p$-values via knockoff technique (see Section 7.2.6 in \cite{ModelXknockoff}).
\end{itemize}

\begin{table*}[!htbp]
\small
\centering
\caption{Algorithmic properties ($\checkmark$-has the given information, $\times$-hasn't the given information)}
\label{algorithm comparison}
\begin{tabular*}{17.5cm}{@{\extracolsep{\fill}}l|ccccccc}
\hline
Properties& FX-Knockoff&MX-Knockoff&DeepPINK&RANK&PC-Knockoff&E-Knockoff (Ours)&\\
\hline
Coefficient-free feature statistics&  $\times$ &$\times$ & $\times$ & $\times$ & $\checkmark$ & $\checkmark$\\
$k$-FWER control& $\times$ &$\times$ & $\times$ & $\times$ & $\times$ & $\checkmark$\\
FDP control&$\times$ & $\times$ &$\times$ & $\times$ & $\times$ & $\checkmark$\\
FDR control& $\checkmark$ &$\checkmark$ & $\checkmark$ & $\checkmark$ & $\checkmark$ & $\checkmark$ \\
Robust analysis&$\times$ & $\times$ &$\times$ & $\checkmark$ & $\times$ & $\checkmark$\\
Power analysis (linear model)& $\times$ &$\times$ & $\times$ & $\checkmark$ & $\checkmark$ & $\checkmark$\\
Power analysis (nonlinear model)& $\times$ &$\times$ & $\times$ & $\times$ &$\checkmark$ & $\checkmark$\\
\hline
\end{tabular*}
\end{table*}

\subsection{Main Contributions}
To address the above issues, this paper proposes a new knockoff filter scheme, called  \emph{Error-based Knockoffs Inference} (E-Knockoff), for controlled feature selection based on the error-based feature statistics.
The main contributions of this paper are summarized as below:

\begin{itemize}

\item \emph{Error-based knockoffs inference}.
Our model integrates the  knockoff features \cite{ModelXknockoff}, the error-based feature statistics and the stepdown procedure \cite{Lehmann_K-FWER_FDP_2005} into a coherent way for FDR, FDP or $k$-FWER control.
The error-based importance measure does not require specifying a regression model and  can be used to calculate the valid $p$-values efficiently.
The stepdown procedure, with the help of new importance statistics, provides the route to control FDP and $k$-FWER, respectively, which is different from the previous knockoffs inference just for FDR control \cite{Knockoff_FDR,ModelXknockoff,LuDeepPINK,Knockoff_RANK,knockoff_ipad,knockoff_model-free}. In particular, it is novel to design the error-based  statistics of feature importance under the knockoff framework, which partially answers the open questions stated  in Sections 7.2.5 and 7.2.6 \cite{ModelXknockoff}.

\item \emph{Theoretical guarantees on $k$-FWER, FDP, and FDR control}.
For the MX-Knockoff framework, statistical foundations on the power and FDR control have been provided in \cite{ModelXknockoff,Knockoff_RANK,knockoff_ipad}, where the power analysis is limited to high-dimensional linear models with both known and unknown covariate distribution. Beyond the linear models in aforementioned literature, we state theoretical justifications on the tight $k$-FWER control and FDP control when the stepdown procedure is employed. In particular, the robustness of $k$-FWER and FDP control can also be assured even for unknown covariate distribution associated with Gaussian graphical model. Additionally, our power analysis holds for general nonlinear models, which is closely related to the open question illustrated in Section 6 \cite{Knockoff_RANK}.   Some empirical evaluations support our theoretical findings.
\end{itemize}

To better illustrate the novelty of current work, we compare it with FX-Knockoff \cite{Knockoff_FDR}, MX-Knockoff \cite{ModelXknockoff}, DeepPINK \cite{LuDeepPINK}, RANK \cite{Knockoff_RANK}, PC-Knockoff \cite{knockoff_model-free} in Table \ref{algorithm comparison} from the lens of  feature statistics, control ability, and asymptotic theory. Table \ref{algorithm comparison} shows  that our approach enjoys theoretical guarantees on robustness and power for FDR, FDP, and $k$-FWER control.





\section{Preliminaries}
This section introduces some necessary backgrounds including the problem setup, the knockoff filter \cite{ModelXknockoff} and the stepdown procedure \cite{Lehmann_K-FWER_FDP_2005}.

\subsection{Problem Statement}
Let $\mathcal{X} \subset \mathbb{R}^p$ be the compact input space and let $\mathcal{Y}\subset \mathbb{R}$ be the output set.
We have $n$ independent identically distributed (i.i.d.) observations $\{(\mathbf{x}_i,Y_i)\}_{i=1}^n$  from the population $(\mathbf{x},Y)$, where $\mathbf{x} = (X_1,\dots,X_p) \in \mathcal{X}$ and $Y \in \mathcal{Y}$. Suppose that the conditional distribution of $Y$ is only relevant with a small subset of $p$ covariates.

The definition of irrelevant features is given in   \citet{ModelXknockoff}.
\begin{definition}\label{irrelevant_group}\cite{ModelXknockoff} A feature $X_j$ is said to be ``irrelevant" if $Y$ is independent of $X_j$ conditionally on $$\mathbf{x}_{-j}:= (X_1, \dots , X_{j-1}, X_{j+1}, \dots , X_p).$$
	For simplicity, we denote it as
\begin{equation*}
Y  \rotatebox[]{90}{$\models$}  X_j | \mathbf{x}_{-j}.
\end{equation*}
\end{definition}
\begin{remark}
If feature $X_j$ is irrelevant according to Definition \ref{irrelevant_group}, it satisfies that  the conditional distribution $Y|\mathbf{x} \overset{d}= Y|\mathbf{x}_{-j}$, where $\overset{d}=$ denotes equality in distribution. That is, the conditional distribution of \ $Y$ remains invariant when removing $X_j$ from  $\mathbf{x}$.
\end{remark}
Let $\mathcal{S}_1 \subset \{1, \dots, p\}$ be the index set with respect to irrelevant features. Naturally,  the index set of true informative features  $\mathcal{S}_0$ is the complement set of $\mathcal{S}_1$, i.e., $\mathcal{S}_0=\mathcal{S}_1^c$. This paper aims to find $\widehat{\mathcal{S}}$, the data dependent estimation of $\mathcal{S}_0$, while controlling $k$-FWER, FDP, or FDR. Recall that
\begin{equation*}
    \text{FDR} = \mathbb{E}[\text{FDP}] = \mathbb{E}\left[\frac{|\widehat{\mathcal{S}} \cap \mathcal{S}_1|}{|\widehat{\mathcal{S}}| \vee 1}\right]
    \end{equation*}
    and
  \begin{equation*}
\text{$k$-FWER} = Prob\{|\widehat{\mathcal{S}} \cap \mathcal{S}_1| \geq k\},
\end{equation*}
where $|\cdot|$ is the cardinality of a set.

\subsection{Model-X Knockoff Framework}\label{KnockoffFilterReview}
Model-X knockoff framework \cite{ModelXknockoff} aims to identify informative features while controlling FDR. The key point is to construct the knockoff copy of $\mathbf{x}$ which looks like real ones without contribution to the response.
\begin{definition}\cite{ModelXknockoff} \label{def-knockoffs}
	Model-X knockoffs for the family of random variables $\mathbf{x}= (X_1,\dots,X_p)$ is a new family of random variables $\tilde{\mathbf{x}} = (\tilde{X}_1,\dots,\tilde{X}_p)$ satisfying
\begin{equation}\label{knockoff_property_1}
\tilde{\mathbf{x}} \rotatebox[]{90}{$\models$} Y |\mathbf{x}
\end{equation}
and
\begin{equation}\label{knockoff_property_2}
(\mathbf{x}, \tilde{\mathbf{x}})_{swap(s)} \overset{d}= (\mathbf{x}, \tilde{\mathbf{x}}),\forall s \subset \{1, \dots, p\}.
\end{equation}
Here,  $(\mathbf{x}, \tilde{\mathbf{x}}) = (X_1,\dots,X_p, \tilde{X}_1,\dots,\tilde{X}_p)$,
and $(\mathbf{x}, \tilde{\mathbf{x}})_{swap(s)}$ is swapping $X_j$ with $\tilde{X}_j$ for all $j \in s$, e.g., when $p = 3$, $(\mathbf{x},\tilde{\mathbf{x}})_{swap(\{2,3\})} = (X_1, \tilde{X}_2,\tilde{X}_3, \tilde{X}_1, X_2,X_3)$.
\end{definition}

\begin{remark}
The properties of knockoff features have been well investigated in   \cite{ModelXknockoff}.
The property \eqref{knockoff_property_1} illustrates that all knockoff features are noise features, and \eqref{knockoff_property_2} assures the similarity between $\mathbf{x}$ and $\tilde{\mathbf{x}}$.
\end{remark}

 \citet{ModelXknockoff} state the construction of knockoffs when $\mathbf{x}$ obeys a known Gaussian graphical model $\mathcal{N}(\mathbf{0}, \mathbf{\Sigma})$, where the covariance matrix $\mathbf{\Sigma}$ is positive definite. Model-X knockoff can construct $\tilde{\mathbf{x}}$ conditionally on $\mathbf{x}$ w.r.t. $\tilde{\mathbf{x}}|\mathbf{x} \overset{d}= \mathcal{N}(\mu,\mathbf{V})$, where
\begin{equation*}\label{Definition-mu}
\mu = \mathbf{x} (I_p- \mathbf{\Sigma^{-1}} \text{diag}\{\mathbf{s}\}),
\end{equation*}
\begin{equation*}\label{Definition-V}
\mathbf{V} = 2\text{diag}\{\mathbf{s}\} - \text{diag}\{\mathbf{s}\}\mathbf{\Sigma^{-1}} \text{diag}\{\mathbf{s}\},
\end{equation*}
and the joined distribution of $(\mathbf{x},\tilde{\mathbf{x}})$ satisfies
\begin{equation*}\label{Distribution-of-joined-features}
(\mathbf{x},\tilde{\mathbf{x}}) \sim \mathcal{N}(\mathbf{0}, \mathbf{G})
\end{equation*}
with
\begin{equation}
\label{Distribution-of-joined-features}\mathbf{G} =
\left(
\begin{array}{cc}
\mathbf{\Sigma} & \mathbf{\Sigma} - \text{diag}\{\mathbf{s}\} \\
\mathbf{\Sigma} - \text{diag}\{\mathbf{s}\}& \mathbf{\Sigma}
\end{array}
\right).
\end{equation}


Some strategies have been provided in \cite{ModelXknockoff} for selecting diagonal matrix $\text{diag}\{\mathbf{s}\}$ .

Given i.i.d. observations $\{(\mathbf{x}_i,Y_i)\}_{i=1}^n$, denote
$$\mathbf{X}= (\mathbf{x}_i)_{i=1}^n\in\mathbb{R}^{n\times p}~~ \mbox{and}~~
\mathbf{y}=(Y_i)_{i=1}^{n}\in\mathbb{R}^{n},$$
 where each $\mathbf{x}_i =(X_{i1},\dots,X_{ip})$.
The knockoff data matrix $\tilde{\mathbf{X}} = (\tilde{\mathbf{x}}_i)_{i=1}^n$ is constructed by row w.r.t. $\tilde{\mathbf{x}}|\mathbf{x}$, where $\tilde{\mathbf{x}}_i$ is the knockoff copy of $\mathbf{x}_i$. The $n \times 2p$ matrix $[\mathbf{X}, \mathbf{\tilde{X}}] $ is obtained by connecting $\mathbf{X}$ and $\tilde{\mathbf{X}}$.
To identify active features, a paired-input filter is trained on $\big([\mathbf{X}, \tilde{\mathbf{X}}],\mathbf{y}\big)$, e.g, Lasso \cite{hastie} in  \cite{ModelXknockoff}. 
Then, each feature (including its knockoff) is assigned with a coefficient-based score, e.g., the absolute value of Lasso coefficient \cite{ModelXknockoff,Knockoff_RANK}.

Let $Z_j$ and $\tilde{Z}_j$ be the score of $X_j$ and $\tilde{X}_j$ respectively. The  importance measure of feature $X_j$ is defined by
 $$W_j:=w_j\big([\mathbf{X},\mathbf{\tilde{X}}],\mathbf{y}\big) = Z_j - \tilde{Z}_j,$$
where  $w_j$ is a model-driven function associated with $[\mathbf{X},\mathbf{\tilde{X}}],\mathbf{y}$.
Typical example of $W_j$ is the \emph{Lasso coefficient difference} (LCD) used in \cite{ModelXknockoff,Knockoff_RANK}.
The distribution of $W_j$ enjoys the following \emph{flip-sign} property.
\begin{proposition}\label{Model-X W_j distribution}\cite{ModelXknockoff}
Assume swapping $X_j$ with $\tilde{X}_j$ has the effect of changing the sign of $W_j$, i.e.,
\begin{equation*}
    w_j\big([\mathbf{X},\mathbf{\tilde{X}}]_{swap(\{j\})},\mathbf{y}\big) = -w_j\big([\mathbf{X},\mathbf{\tilde{X}}],\mathbf{y}\big).
\end{equation*}
Then, each $W_j$ associated with irrelevant feature is independent and symmetrically distributed.
\end{proposition}
The above property is helpful to obtain the theoretical guarantee on the FDR control.
\begin{lemma}
\label{FDR_control}
\cite{ModelXknockoff} If \ $W_j$ is independent and symmetrically distributed for each $j \in \mathcal{S}_1$. For any given target FDR level $q \in (0,1)$, let
\begin{equation}\label{feature_threshold_t}
\tau = \min \bigg\{ \tau>0: \frac{1 + |\{j:W_j \leq -\tau\}| }{|\{j: W_j\geq \tau\}| \vee 1} \leq q\bigg\}.
\end{equation}
Then the procedure selecting the variables $\widehat{\mathcal{S}} = \{j:W_j \geq \tau\}$ can control
$\text{FDR}\leq q$.
\end{lemma}

Usually, the FDR control under model-X knockoff framework depends heavily on the coefficient difference derived from Lasso \cite{ModelXknockoff,Knockoff_RANK},  group Lasso \cite{Knockoff_group_zoom},  and paired-input deep neural networks \cite{LuDeepPINK}. In many applications involving complex function relationships, this coefficient-based property may hinder the flexibility and accuracy of model-X knockoff framework.



\subsection{Stepdown Procedure}
Denote  $\mathcal{P}_j$ as  the $p$-value associated with the significance of feature $X_j, j=1,\dots, p$.
Let $\mathcal{P}_{k_j}$, $j=1,\dots, p$, be the $p$-values with $\mathcal{P}_{k_1} \leq  \dots \leq \mathcal{P}_{k_p}$ and  let $\alpha_j,j=1,\dots, p$, be the significance threshold values with $\alpha_1 \leq \dots \leq \alpha_p$. Naturally, the first $m$ features with lower $p$-values are selected as informative variables $\mathcal{\widehat{S}} = \{k_1, \dots, k_m\}$, where
\begin{equation*}
    m = \max\{M: \mathcal{P}_{k_j} \leq \alpha_j, \forall j \leq M\}.
\end{equation*}
\citet{Lehmann_K-FWER_FDP_2005} have stated the following results about $k$-FWER  and FDP control.
\begin{lemma}\label{kFWER-control}\cite{Lehmann_K-FWER_FDP_2005}
For any given $\alpha \in (0,1)$ and $k = 1, \dots ,p$, the stepdown procedure with
\begin{equation*}
\alpha_j=\left\{
\begin{aligned}
&\frac{k\alpha}{p},& &j \leq k\\
&\frac{k\alpha}{p + k - j}, & &j > k
\end{aligned}
\right.
\end{equation*}
can control $\text{$k$-FWER} \leq \alpha$.
\end{lemma}


\begin{lemma}\label{FDP control}\cite{Lehmann_K-FWER_FDP_2005}
For any given $\alpha,q \in (0,1)$, if the $p$-value of any irrelevant feature is independent of the $p$-values of informative features,
 the stepdown procedure with $$\alpha_j = \frac{(\lfloor qj \rfloor + 1)\alpha}{p+\lfloor qj \rfloor + 1 - j}$$ satisfies $Prob\{\text{FDP} > q\} \leq \alpha$.
\end{lemma}

\section{Error-based Knockoff Inference}
 This paper exploits the idea of ``feature replacing" for controlled feature selection, i.e., replacing a feature with its knockoff and see whether there is a significant difference in the estimation error or not. We first assume the distribution of $\mathbf{x}$ is known as prior and propose an error-based feature statistic for $k$-FWER, FDP, or FDR control. Then, we extend the theoretical result to a more general setting where the distribution of $\mathbf{x}$ is unknown. Finally, the power analysis is stated for the proposed approach.

\subsection{Error-based Feature Importance}
To construct the error-based feature importance $W_j,j=1,\dots,p$, we need to divide $n$ samples  into two disjointed parts: $(\mathbf{X}^{*},\mathbf{y}^{*})$ and $(\mathbf{X}',\mathbf{y}')$, containing $n_1$ and $n_2$ samples, respectively.

Let $f$ be the regression estimator trained on $(\mathbf{X}^{*},\mathbf{y}^{*})$ and let $\tilde{\mathbf{X}}'$ be the knockoff copy of $\mathbf{X}'$. Denote $\big((\mathbf{x}_i', \tilde{\mathbf{x}}_i'),Y_i'\big)$ as the $i$-th column of $([\mathbf{X}', \tilde{\mathbf{X}}'],\mathbf{y}')$. Given an estimator $f$ and random sample $(\mathbf{x},Y)\in\mathcal{X}\times\mathcal{Y}$, define the error-based random variable
\begin{equation}\label{error-random1}
\xi:=\xi(\mathbf{x},Y) = |f(\mathbf{x}) - Y|
\end{equation}
and
\begin{equation}\label{error-random2}
\xi^j:=\xi^j(\mathbf{x},Y) = |f\big(\mathcal{R}_j(\mathbf{x})\big) - Y|,
\end{equation}
 where
\begin{equation*}\label{R-j}
\mathcal{R}_j(\mathbf{x}) = (X_1,\dots,X_{j-1},\tilde{X}_j,X_{j+1}, \dots,X_p).
\end{equation*}
For observation $(\mathbf{x}'_{i},Y'_{i})$ associated with $(\mathbf{X}',\mathbf{y}')$,  we define
$$
\xi_i= \left| f(\mathbf{x}'_{i}) - Y'_{i} \right|
~~\mbox{and}~~
\xi_i^j= \left| f\big(\mathcal{R}_j(\mathbf{x}_{i}')\big) - Y'_{i} \right|.
$$

The feature importance can be measured by the \emph{error difference}  between $\xi_i^j$ and $\xi_i$, i.e., $$T_i^j := \xi_i^j - \xi_i.$$
 The error-based feature importance can be characterized by
\begin{equation}\label{feature importance}
W_j: = \frac{1}{n_2}\Big(\sum\limits_{i = 1}^{n_2}\rm{I}_{\{\it{T_i^j} > 0\}}\Big) - 0.5,
\end{equation}
where indicator function $\rm{I}_{\{\it{A}\}}=1$ if $A$ is true and 0 otherwise.

The basic properties of $W_j$ are stated as below, which are obtained from the definition of knockoff features (e.g., Definition \ref{def-knockoffs}) and the \emph{flip-sign} property of MX-Knockoff feature importance (e.g., Proposition  \ref{Model-X W_j distribution}). The corresponding proof can be found in \emph{Supplementary Material C}.
\begin{proposition}\label{independnece}
For each $j \in \mathcal{S}_1$,  $W_j$ defined in (\ref{feature importance}) is independent and  symmetrically distributed around zero, and satisfies
$n_2(W_j + 0.5) \sim \mathcal{B}(n_2, 0.5), \forall j \in \mathcal{S}_1$.
\end{proposition}
\begin{remark}
The first conclusion of Theorem \ref{independnece} differs from Proposition \ref{Model-X W_j distribution} ( See also Lemma 3.3 in \cite{ModelXknockoff}) in that, we remove the assumption on feature importance via replacing strategy. Since the error-based importance measure has no requirement on the structure of learning machine, it may be much flexible for applications.  The second conclusion reveals the distribution information of the proposed irrelevant feature's importance, which  gives us the opportunity to realize $k$-FWER control and FDP control via combining the knockoff technique with the stepdown procedure.
\end{remark}

Combining Lemma \ref{FDR_control} and Proposition \ref{independnece} yields the following result for FDR control.
\begin{theorem}\label{ours_FDR_control}
For any given target FDR level $q \in (0,1)$, the error-based knockoff procedure, with feature importance \eqref{feature importance} and knockoff threshold \eqref{feature_threshold_t},  satisfies  $\text{FDR} \leq q$.
\end{theorem}

 Assume  the null-hypothesis that the feature is irrelevant. Let $M_j := \max\{n_2(W_j + 0.5), n_2(0.5 - W_j)\}$. The $p$-values are defined as
\begin{equation}\label{p-value_gen}
\mathcal{P}_j := 2\sum\limits_{i =M_j}^{n_2} C(n_2,i) \frac{1}{2^{n_2}},  j=1,\dots,p
\end{equation}
are used to evaluate the feature significance, where $C(n_2,m)$ is the combinatorial number.

The following theoretical results on $k$-FWER and FDP control can be established by combining Proposition \ref{independnece} with Lemmas \ref{kFWER-control} and \ref{FDP control}.
\begin{theorem}
\label{Ours_k-FWER_control}
For any given $\alpha \in (0,1)$,  the stepdown procedure,  constructed in Lemma \ref{kFWER-control} and associated with knockoff-based  $p$-values \eqref{p-value_gen}, satisfies $\text{$k$-\text{FWER}} \leq \alpha$.
\end{theorem}
\begin{theorem}
\label{Ours_FDP_control}
For any given $q,\alpha \in (0,1)$,  the FDP of $\widehat{\mathcal{S}}$, associated with the stepdown procedure in Lemma \ref{FDP control} and $p$-values in \eqref{p-value_gen}, satisfies $Prob\{\text{FDP} > q\} \leq \alpha$.
\end{theorem}

\begin{remark}
Different from the previous knockoff filters relied  on the \textbf{coefficient difference} \cite{Knockoff_FDR,ModelXknockoff,LuDeepPINK,Knockoff_high_dim}, the current knockoff procedure rooted in the \textbf{error difference}. The error-based knockoff strategy is model-free (no structure restriction on estimator $f$), and gives us the opportunity to tackle FDR, FDP, and $k$-FWER control.
\end{remark}

\begin{algorithm}[t]
\caption{Construct feature importance statistic $W_j$}
\label{W_algorithm}
\textbf{Input}: Data $(\mathbf{X}',\mathbf{y}')$, trained filter $f$, feature index $j$\\
\textbf{Output}: Feature importance statistic $W_j$\\
\begin{algorithmic}[1] 
\STATE Construct $\mathbf{\tilde{X}}'$, i.e., the knockoff copy of $\mathbf{X}'$.
\FOR{$i= 1,\dots,n_2$}
\STATE Obtain $\mathcal{R}_j(\mathbf{x}'_i)$ by replacing $j$-th feature in $\mathbf{x}'_{i}$ with its knockoff copy.
\STATE $T_i^j \leftarrow \left| f\big(\mathcal{R}_j(\mathbf{x}'_i)\big) - Y'_i \right| -  |f(\mathbf{x}'_{i}) - Y'_{i} |$
\ENDFOR
\STATE $W_j \leftarrow \frac{1}{n_2}\left(\sum\limits_{i = 1}^{n_2}\rm{I}_{\{\it{T_i^j} > 0\}}\right) - 0.5.$
\RETURN $W_j $
\end{algorithmic}
\end{algorithm}

\subsection{Robustness Analysis}
This section further  establishes the asymptotic properties of $k$-FWER control and FDP control when the distribution of $\mathbf{x}$ is characterized by some unknown Gaussian graphical model, i.e., $\mathbf{x} \sim \mathcal{N}(\mathbf{0},\mathbf{\Sigma})$. All proofs of this section have been provided in \emph{Supplementary Material C}.

Let $\mathbf{\widehat{\Sigma}}$ be the empirical estimation of covariance matrix obtained by $(\mathbf{X}^*,\mathbf{y}^*)$.
To ease the presentation, for any notation $\mathbf{A}$ associated with the unknown covariance matrix $\mathbf{\Sigma}$, the notation $\widehat{\mathbf{A}}$ stand for its empirical estimation constructed via $\widehat{\mathbf{\Sigma}}$. Inspired from \citet{Knockoff_RANK}, we introduce the following conditions for our robustness analysis.

The following condition on density function is required, which holds true for bounded regression problem with  Gaussian noise assumption \cite{Lasso,group_lasso,group_lr,Ridge}.
\begin{condition}\label{bounded density}
Let $\eta(Y|\mathbf{x})$ be the probability density function of $Y$ conditioned on $\mathbf{x}$. There holds
$\max\limits_{(\mathbf{x},Y)}\eta(Y|\mathbf{x}) \leq C_1$
for some constant $C_1$.
\end{condition}

Without loss of generality, assume the covariance matrix $\mathbf{G}$ defined in \eqref{Distribution-of-joined-features} to be positive definite \cite{Knockoff_RANK}. The following condition is used to characterize the relationship between $\mathbf{G}$ and its empirical estimation $\widehat{\mathbf{G}}$, and to rule out some extreme case of these matrixs,  e.g., $\lambda_{max}(\mathbf{G}) = \infty$. 
Similar condition has been used  in \cite{Knockoff_RANK} for robust analysis.

\begin{condition}\label{covariancematrix}
Let $\lambda_{min}(\cdot)$ and $\lambda_{max}(\cdot)$ be  the  minimum and the maximum matrix eigenvalues, respectively.
There exist some  positive sequence $a_{n_1},b_{n_1}$ satisfying $a_{n_1}\rightarrow 0,b_{n_1}\rightarrow 0$ as $n_1\rightarrow\infty$, and a positive constant $C_2$ such that
\begin{equation*}
\|\widehat{\mathbf{G}} - \mathbf{G}\|_2 \leq a_{n_1}
\end{equation*}
and
\begin{equation*}
\begin{split}
    \frac{1}{C_2} \leq & \min \bigg\{ \lambda_{\min}(\mathbf{G}),  \lambda_{\min}(\mathbf{\widehat{G}}) \bigg\} \\
\leq & \max \bigg\{ \lambda_{\max}(\mathbf{G}),  \lambda_{\max}(\mathbf{\widehat{G}}) \bigg\} \leq C_2
\end{split}
\end{equation*}
with  probability at least $1 - p^{-\frac{1}{b_{n_1}}}$.
\end{condition}

\newcommand{\tabincell}[2]{\begin{tabular}{@{}#1@{}}#2\end{tabular}}
\begin{table*}[!t]
\renewcommand\arraystretch{1.3}
\centering
\caption{Results on the simulated data for controlled feature selection  (different dimension $p$)}\label{Simulation_result_addative}
\resizebox{\textwidth}{!}{
\begin{tabular}{c|ccc|ccc|ccc|ccc|ccc}
\hline
\multirow{2}*{\tabincell{c}{$p$}}&\multicolumn{3}{c|}{E-Knockoff($k$-FWER)}&\multicolumn{3}{c|}{E-Knockoff(FDP)}&\multicolumn{3}{c|}{E-Knockoff(FDR)}&
\multicolumn{3}{c|}{MX-Knockoff}&\multicolumn{3}{c}{DeepPINK}\\
&$\text{FDP}_{\max}$&FDR&Power&$\text{FDP}_{\max}$&FDR&Power&$\text{FDP}_{\max}$&FDR&Power&$\text{FDP}_{\max}$&FDR&Power&$\text{FDP}_{\max}$&FDR&Power\\
\hline
50  & 0.03 & 0.01 & 1.00 & 0.12 & 0.03 & 1.00 & 0.35 & 0.19 & 1.00 & 0.33 & 0.20 & 1.00 & 0.32 & 0.20 & 1.00\\
100 & 0.03 & 0.00 & 0.97 & 0.17 & 0.04 & 1.00 & 0.40 & 0.19 & 1.00 & 0.45 & 0.19 & 1.00 & 0.52 & 0.17 & 1.00\\
200 & 0.07 & 0.00 & 0.95 & 0.14 & 0.04 & 0.99 & 0.48 & 0.20 & 1.00 & 0.45 & 0.20 & 1.00 & 0.36 & 0.16 & 1.00\\
400 & 0.04 & 0.01 & 0.91 & 0.13 & 0.04 & 0.98 & 0.43 & 0.19 & 1.00 & 0.39 & 0.19 & 1.00 & 0.50 & 0.22 & 1.00\\
800 & 0.07 & 0.01 & 0.63 & 0.13 & 0.04 & 0.83 & 0.36 & 0.19 & 0.95 & 0.40 & 0.20 & 1.00 & 0.33 & 0.18 & 1.00\\
1200& 0.06 & 0.00 & 0.60 & 0.08 & 0.03 & 0.79 & 0.42 & 0.16 & 0.93 & 0.40 & 0.16 & 1.00 & 0.38 & 0.19 & 1.00\\
1600& 0.07 & 0.01 & 0.59 & 0.17 & 0.04 & 0.79 & 0.40 & 0.18 & 0.94 & 0.35 & 0.19 & 1.00 & 0.44 & 0.18 & 1.00\\
2000& 0.07 & 0.01 & 0.57 & 0.13 & 0.03 & 0.77 & 0.52 & 0.19 & 0.92 & 0.41 & 0.16 & 1.00 & 0.44 & 0.24 & 0.94\\
\hline
\end{tabular}}
\end{table*}

Let $\tilde{\mathbf{x}}^{\widehat{\mathbf\Sigma}}$ be the knockoff feature based on the distribution $\mathcal{N}(\mathbf{0},\widehat{\mathbf\Sigma})$. For feasibility, denote $\eta_{\mathbf{\Sigma}}$ and $\eta_{\mathbf{\widehat{\Sigma}}}$ be the distribution density function of $(\mathbf{x}, \tilde{\mathbf{x}})$ and $(\mathbf{x}, \tilde{\mathbf{x}}^{\widehat{\mathbf\Sigma}})$ respectivly. The relationship between $\eta_{\mathbf{\Sigma}}$ and $\eta_{\mathbf{\widehat{\Sigma}}}$ is described as below.
\begin{lemma}\label{density}
Under Condition \ref{covariancematrix},   there holds
\begin{equation*}
|\eta_{\mathbf{\Sigma}}(\mathbf{x},\mathbf{\tilde{x}}) - \eta_{\mathbf{\widehat{\Sigma}}}(\mathbf{x},\mathbf{\tilde{x}})| \leq O(a_{n_1}), \forall (\mathbf{x},\mathbf{\tilde{x}})\in\mathcal{X}^2
\end{equation*}
with  probability at least $1 - p^{-\frac{1}{b_{n_1}}}$.
\end{lemma}

It is a position to present the main results on robustness analysis for controlled variable selection.

\begin{theorem}\label{robust k-FWER control}
Let Conditions \ref{bounded density} and \ref{covariancematrix} be true. For any given $\alpha \in (0,1), k = 1, \dots, p$ and $n_2 \in \mathbb{R}$, the feature seleciton procedure described in Theorem \ref{Ours_k-FWER_control} satisfies
\begin{equation*}
      \widehat{\text{$k$-FWER}} \leq \alpha + O(p^{-\frac{1}{b_{n_1}}})+ O(a_{n_1}).
\end{equation*}
\end{theorem}

\begin{theorem}\label{robust FDP control}
Let Conditions \ref{bounded density} and \ref{covariancematrix} be true. For any given $q,\alpha \in (0,1)$, $n_2 \in \mathbb{R}$, the feature selection procedure described in Theorem \ref{Ours_FDP_control} satisfies
\begin{equation*}
    Prob\{\widehat{\text{FDP}} > q \}  \leq \alpha + O(p^{-\frac{1}{b_{n_1}}}) + O(a_{n_1}).
\end{equation*}
\end{theorem}

\begin{table}[t]
\renewcommand\arraystretch{1.3}
\centering
\caption{Maximum number of false discoveries in 50 trials}\label{Simulation_result_k-FWER}
\resizebox{8.0cm}{!}{
\begin{tabular}{c|cccccccc}
\hline
method&50&100&200&400&800&1200&1600&2000\\
\hline
E-Knockoff($k$-FWER)&1&1&2&1&1&1&1&1\\
\hline
\end{tabular}}
\end{table}
\begin{remark}
Theorems \ref{robust k-FWER control} and \ref{robust FDP control} guarantee the robustness of our error-based knockoff inference under mild conditions. Here, we omit the robust analysis for FDR control since it can be derived directly from \cite{Knockoff_RANK}. To the best of  our knowledge, there is no robust analysis for FDP and $k$-FWER control under the knockoff filtering framework.
\end{remark}

\begin{figure*}[!t]
\centering
\includegraphics[scale=0.49]{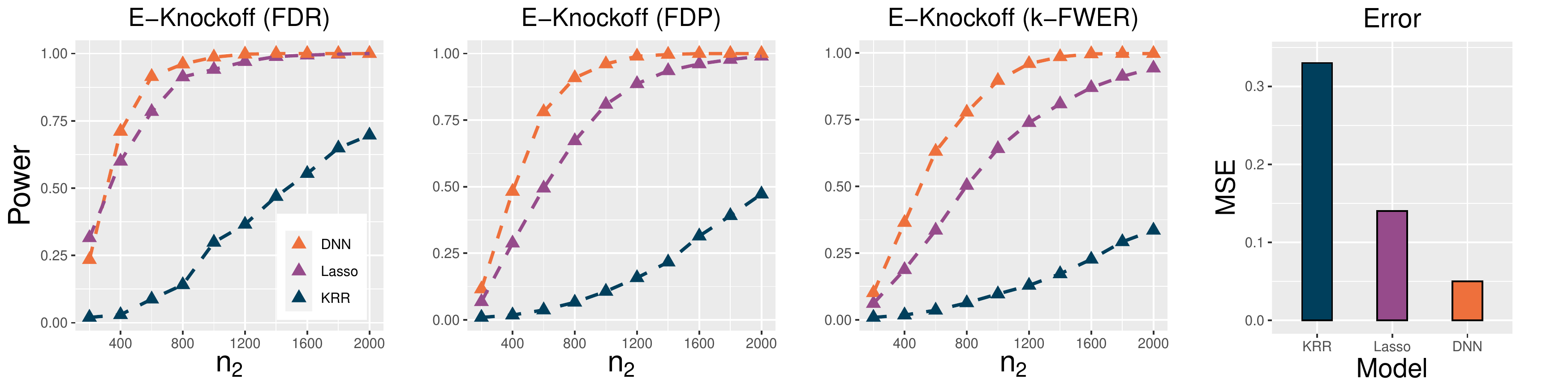}
\caption{Power analysis (Power Vs. $n_2$)  on the simulated data  for different learning models}
\label{Power_results}
\end{figure*}

\subsection{Power Analysis}
The following restriction on the predictor $f$ is involved for the power property of our error-based knockoff inference.
\begin{condition}\label{relevant bound}
For the error-based random variables $\xi$ in \eqref{error-random1} and $\xi^j$ in \eqref{error-random2}, there holds
\begin{equation*}\label{informative group bound}
   Prob\{ \xi^j > \xi \} > Prob\{\xi^j < \xi \},  \forall j\in\mathcal{S}_0.
\end{equation*}
\end{condition}
Condition \ref{relevant bound} ensures that  the predictor $f$ would have degraded performance when replacing an informative feature with a knockoff feature. Recall that
the construction procedure of knockoffs   is independent of the response $Y$, see e.g.,  \citet{ModelXknockoff,DeepKnockoffs,knockoffgan}.
Therefore, the current restriction on $f$ is mild  since less information is more likely to result in additional prediction loss.

The central limit theorem assures that the variance of proposed feature importance in \eqref{feature importance} would converge to zero. Thus, we get the following result.
\begin{theorem}\label{Power guarantee}
Under Condition \ref{relevant bound},  the feature selection procedures described in Theorem \ref{ours_FDR_control}-\ref{Ours_FDP_control} satisfy
\begin{equation*}
     \text{Power} := \mathbb{E}\left[\frac{|\widehat{\mathcal{S}} \cap \mathcal{S}_0|}{|\mathcal{S}_0|} \right] \rightarrow  1 ~~\mbox{as}~~n_2 \rightarrow \infty.
\end{equation*}
\end{theorem}

\begin{remark}
Theorem \ref{Power guarantee} demonstrates that the power of proposed procedures depends on the sample size $n_2$ of $(\mathbf{X}', \mathbf{y}')$. Theorems \ref{robust k-FWER control}-\ref{Power guarantee} imply that, for our error-based knockoff inference, there is a tradeoff between $n_1$ (associated with getting the predictor $f$ and covarince matrix  $\mathbf{\widehat{\Sigma}}$) and $n_2$ (related to  generate knockoffs and error-based feature statistic $W_j$).
\end{remark}

\section{Experimental Analysis}
This section states empirical evaluations of our error-based knockoff inference on both synthetic data and HIV dataset \cite{hivdataset} to valid our theoretical claims about controlled feature selection and power analysis. The detailed experiment settings and some additional experiments are provided in \emph{Supplementary Material  D}.

\subsection{Simulated Data Evaluation}
Inspired by \cite{LuDeepPINK}, we draw $\mathbf{x}$ independently from $\mathcal{N}(\mathbf{0}, \mathbf{\Sigma})$, where $\mathbf{\Sigma}^{-1} = (0.5^{|j-k|})_{1 \leq j,k\leq p}$.  Then, we simulate the response from single index model:
\begin{equation*}
Y = g(\mathbf{x}\beta) + \epsilon,~~  \epsilon \sim \mathcal{N}(0,0.01)
\end{equation*}
where the linkage function
$$g(a) = \sqrt{|a|} + a + a^2 + \text{sin}(a) + \text{arctan}(a), \forall a \in \mathbb{R},$$ $\beta = (\beta_1, \dots, \beta_p)^T$  satisfying $ \beta_j=0, \forall j \in \mathcal{S}_1$ and $\beta_j= 1/|\mathcal{S}_0|$ otherwise.
Here, the sample size $n = 2000$ and  the number of features $p \in\{50,100,200,400,800,1200,1600,2000\}$ with $|\mathcal{S}_0| = 30$ \cite{LuDeepPINK}.

 This paper employes the coefficient-based model-X knockoff \cite{ModelXknockoff} and  DeepPink \cite{LuDeepPINK} as the baselines. We set the target FDR level $q = 0.2$ for all FDR controlled methods, set $q = 0.2$ and $\alpha = 0.2$ for FDP control version of E-Knockoff (\emph{E-Knockoff (FDP)}), and  set $k = 2$ and $\alpha = 0.1$ for $k$-FWER control version of E-Knockoff (\emph{E-Knockoff ($k$-FWER)}). The feature importance is measured by the  coefficient difference associated with Lasso for MX-Knockoff \cite{ModelXknockoff} and associated with paired-input DNNs for DeepPink  \cite{LuDeepPINK}. We use Lasso as the base estimator of our E-Knockoff inference with $n_1 = n_2 = 1000$.  Table \ref{Simulation_result_addative} summaries the estimation of FDR, Power and the maximum value of FDP ($\text{FDP}_{\max}$) with 50 repetitions. In addition, Table \ref{Simulation_result_k-FWER} reports the max number of false discoveries for E-Knockoff ($k$-FWER)  in these trials.
These experimental results show that our error-based knockoff inference can reach the FDR control, FDP control, and  $k$-FWER control flexibily, while  MX-Knockoff and DeepPINK just can control the FDR. Meanwhile, E-Knockoff (FDP) and E-Knockoff ($k$-FWER) also enjoy the promising selection accuracy in almost all settings.The results of Table \ref{Simulation_result_addative} also verify the tradeoff between accuracy and power discussed in \cite{KFWER_FDP_2004,Lehmann_K-FWER_FDP_2005,review_2008}.

To  verify the model-free property and power ability of our approach, we provide an experiment to illusrate the influence of $n_2$ and $f$ on selection results. We set $p=800, n_1 = 1000$ and select $n_2$  from $\{200,400,600,800,\dots, 2000\}$. Three classic learning machines are used to get $f$ including Deep neural networks (DNN) \cite{HINTONdnn}, Lasso \cite{Lasso} and Kernel ridge regression (KRR) \cite{Ridge}. Experimental results of power and mean square error (MSE) are displaced in Figure \ref{Power_results}  after repeating the each experiment 30 times. Full simulated results are presented in \emph{Supplementary Material D}.
It can be observed that a powerful selection result can be made with the increase of $n_2$, which supports our conclusion in Theorem \ref{Power guarantee}. Also, the result implies that a better-trained filter can select true active features with less samples.

\begin{figure}[t]
	\centering
	\includegraphics[scale=0.52]{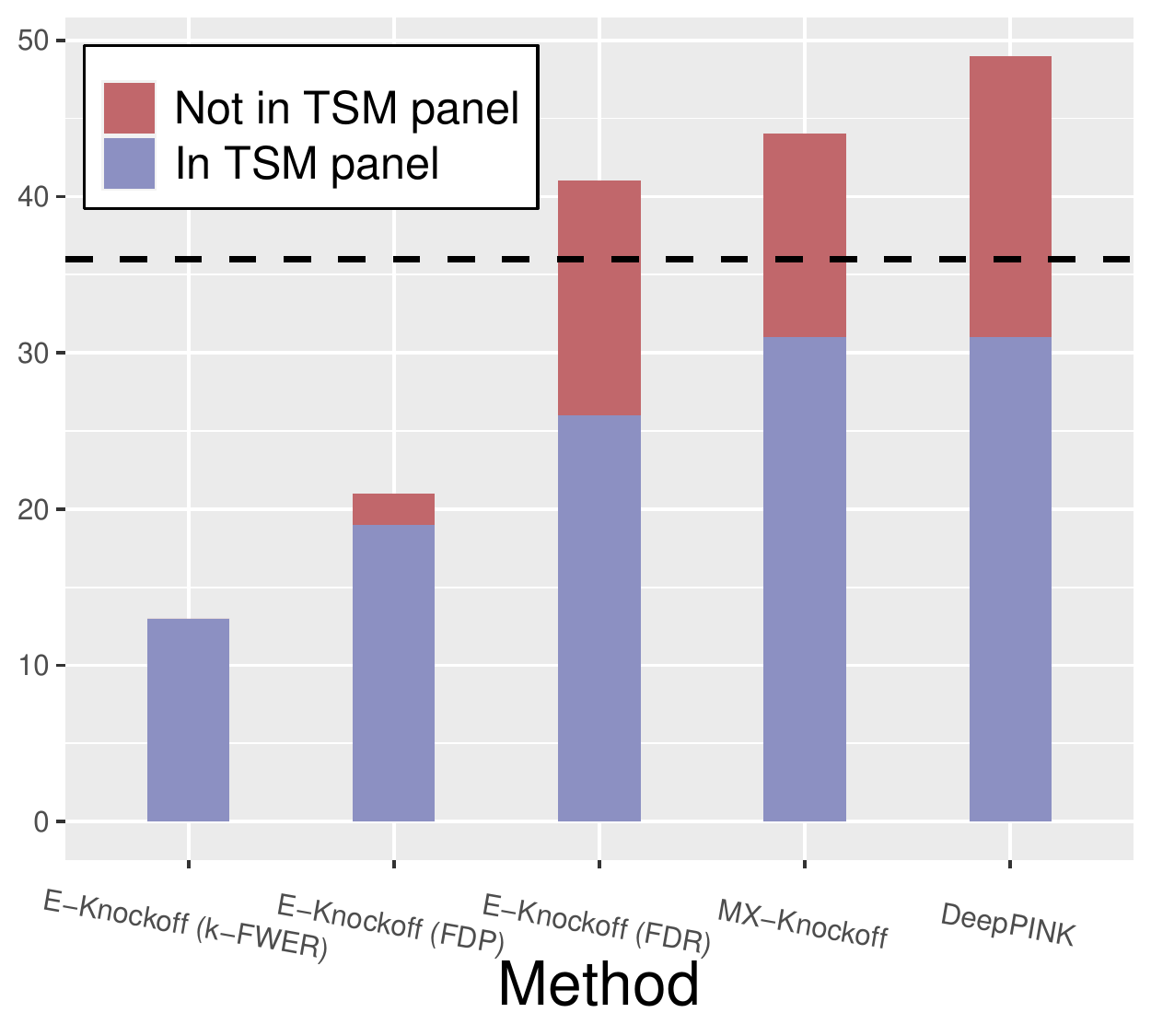}
	\caption{Results on the HIV-1 drug resistance dataset. For FPV drug class, we show the number of mutation positions for PI identified by different knockoff filters. The color indicates whether or not the selected position appears in the TSM panel, and the horizontal line shows the total number of positions on the TSM panel.}
	\label{HIV_results}
\end{figure}

\subsection{Real Data Evaluation}
We next apply E-Knockoff  to identify key mutations of HIV associated with the drug resistance \cite{hivdataset}.
The HIV-1 dataset consists of the data of the drug resistance level, mutations, and the treatment-selected mutations (TSM) associated with drug resistance. For each drug, the response $Y$ is the log-transformed drug resistance level, and the $j$-th feature of argument $\mathbf{x}$ indicates the presence or absence of the $j$-th mutation \cite{LuDeepPINK,multi-environment}.
 Figure \ref{HIV_results} summarizes the experimental results related to FPV drugs resistance (with 1809 samples and 224 dimensions), and \emph{Supplementary Material D} reports the results of other drugs. Here, we use Lasso as the base estimator for E-Knockoff inference ($n_1 = \frac{n}{3}, n_2 = \frac{2n}{3}$). We set  $k = 2, \alpha = 0.1$ for E-Knockoff ($k$-FWER), $q = 0.2, \alpha = 0.2$ for E-Knockoff (FDP), and $q = 0.2$ for E-Knockoff (FDR), MX-Knockoff, and DeepPINK.
Empirical results demonstrate  that the error-based knockoff inference can usually control the false discovery  efficiently.

\section{Conclusion}
To improve the adaptivity and flexibility of the model-X knockoff framework, this paper proposes a new error-based knockoff inference method for controlled feature selection. We establish the statistical asymptotic analysis and power analysis of the proposed approach. Empirical evaluations demonstrate the competitive performance of the proposed procedure  on simulated and real data, which support our research motivation and theoretical findings. In the future, it is interesting to extend the current work for multi-environment controlled feature selection \cite{multi-environment}.

\section{Acknowledgements}
This work was supported by National Natural Science Foundation of China under Grant Nos. 12071166, 62076041, 61702057, 61806027, 61972188, 62106191 and by the Fundamental  Research  Funds  for the Central Universities of China under Grant 2662020LXQD002. We are grateful to the anonymous AAAI reviewers for their constructive comments.

\bibliography{ref}

\section{Supplementary Material A}
To improve the readability,  we summarize the main notations of this paper  in Table \ref{notations}.
\begin{table*}[t]
\renewcommand\arraystretch{1.3}
\centering
\caption{Notations}\label{notations}
\begin{tabular*}{17cm}{l|l}
\Xhline{1.1pt}
Notations&Descriptions\\
\Xhline{1.1pt}

$\mathcal{X}, \mathcal{Y}$&the compact input space and output space, respectively\\
$\mathbf{x}, Y$& the random variables taking values from $\mathcal{X}$ and $\mathcal{Y}$, respectively\\
$n$& sample size\\
$p$& the number of features in $\mathbf{x}$\\
$\mathbf{X}, \mathbf{y} $& the data matrix and data vector assembled by $n$ i.i.d. pairs of $(\mathbf{x},Y)$\\
$\mathbf{X}^{*},\mathbf{y}^{*}$& the data matrix for training the predictor $f$ and covarince matrix $\widehat{\mathbf{\Sigma}}$\\
$\mathbf{X}',\mathbf{y}'$&  the data matrix and data vector related to generating knockoffs and error-based feature statistic\\
$\tilde{\mathbf{x}},\tilde{\mathbf{X}}'$& the knockoff copies of $\mathbf{x}$ and $\mathbf{X}'$, respectively\\
$\mathcal{S}_0,\mathcal{S}_1$& the set of index for informative features and irrelevant features, respectively\\
$\widehat{\mathcal{S}}$ & the data dependent estimation of $\mathcal{S}_0$\\
$W_j,\mathcal{P}_j$ & the knockoff based feature importance and $p$-value for feature $X_j$, respectively\\
$\mathbf{\Sigma}$ & the true covariance matrix of $\mathbf{x}$\\
$\mu,\mathbf{V}$& the mean and covariance matrix of the conditional distribution $\tilde{\mathbf{x}}|\mathbf{x}$, respectively\\
$\mathbf{G}$ & the covariance matrix of joined random variables $(\mathbf{x},\tilde{\mathbf{x}})$\\
\Xhline{1.1pt}
\end{tabular*}
\end{table*}

\section{Supplementary Material B}
\subsection{B.1  Detail experiment settings of Figure 1}
We randomly generated 800 datasets $\{(\mathbf{x}_i^m,Y_i^m)\}_{i=1}^{2000}, m = 1, \dots, 800$, where each dataset contains 2000 i.i.d. observations with 370 irrelevant features and 30 informative features. For each observation $(\mathbf{x}_i^m,Y_i^m)$, the covariate $\mathbf{x}_i^m$ is drawn from $\mathcal{N}(\mathbf{0}, \mathbf{\Sigma})$ with $\mathbf{\Sigma}^{-1} = (0.5^{|j-k|})_{1 \leq j,k\leq 400}$ and the response $y_i^m$ is generated from single index model:
\begin{equation*}
Y_i^m = g(\mathbf{x}_i^m\beta) + \epsilon.
\end{equation*}
Here the linkage function
$$g(a) = \sqrt{|a|} + a + a^2 + \text{sin}(a) + \text{arctan}(a), ~~\epsilon \sim \mathcal{N}(0,0.01)$$
with $ a \in \mathbb{R}, \beta = (\beta_1, \dots, \beta_{400})^T$ satisfying $\beta_j = 1/|\mathcal{S}_0|, \forall j \in \{1, \dots, 30\}$ and $\beta_j = 0$ otherwise.

Figure 1 displays the histogram of FDP over 800 datasets based on the MX-Knockoff filter \cite{ModelXknockoff}, which demonstrates the control of FDR does not assure the control of FDP.

\subsection{B.2  Relationship between FDR  and FDP }
The connection between FDR control and FDP control has been well discussed in \cite{KFWER_FDP_2004,Lehmann_K-FWER_FDP_2005,FDP_2006,FDRisnotFDP}. 

\citet{FDP_2006} proved that if a method controls the FDR at a target level $q$, then
\begin{equation*}
    Prob\Big\{\text{FDP} > \frac{q}\alpha\Big\} \leq \alpha.
\end{equation*}
Obviously, this is result is not tight enough, because the radio $q/\alpha$ can be quite large for very small $\alpha$  \cite{FDP_2006,KFWER_FDP_2004,Lehmann_K-FWER_FDP_2005}.  Moreover, if one insists on controlling FDP via
some FDR control methods, the target FDR level may be too restrictive to produce a powerful selection result, e.g., under the experiment settings of \textbf{B.1} with target level $0.02$, only two informative features are selected by MX-Knockoff ($\alpha = 0.1$, $q/\alpha = 0.2$ and $q = 0.02$).

On the other hand, if a method ensures $Prob\{\text{FDP} > q\} \leq \alpha$, \citet{Lehmann_K-FWER_FDP_2005} established that
\begin{equation*}
    \text{FDR} = \mathbb{E}(\text{FDP}) \leq q+(1-q)\alpha.
\end{equation*}

\section{Supplementary Material C}
Before providing the proofs of Theorems 4-6, we first establish  stepping stones including Propositions 2-3 and Lemmas 4-5.

The following lemma  (See also the proof of Lemma 1 in  \cite{ModelXknockoff}) is key to prove Proposition 2.
\setcounter{lemma}{4}
\begin{lemma}\label{exchangability}
\cite{ModelXknockoff} For any subset $s \subset \mathcal{S}_1$, there holds
\begin{equation*}
\big((\mathbf{x}, \mathbf{\tilde{x}}), Y\big) \overset{d}= \big((\mathbf{x}, \mathbf{\tilde{x}})_{swap(s)}, Y\big).
\end{equation*}
\end{lemma}

\begin{proproof2} The proof proceeds similarly to Lemma 2 in Section 3.2 \cite{ModelXknockoff}.

Recall that the error difference $T_i$ depends on the response, the original input and its knockoffs, i.e.,
\begin{equation*}
T_i = (T_i^1, \dots, T_i^p):= t\big((\mathbf{x}'_i,\tilde{\mathbf{x}}'_i), Y'_i\big)
\end{equation*}
for some model-based function $t(\cdot,\cdot)$.
Let $\epsilon = (\epsilon_1, \dots, \epsilon_p)$ be a sequence of independent variables such that $\epsilon_j = \pm 1$ with probability $\frac{1}{2}$ if $j \in \mathcal{S}_1$, and $\epsilon_j = 1$ otherwise.
To prove the claim, it will suffice to establish that
\begin{equation*}
    T_i \overset{d}= \epsilon \odot T_i,
\end{equation*}
where $\odot$ denotes the point-wise multiplication of vectors $\epsilon$ and $T_i$.

Let $s = \{j: \epsilon_j = -1\}$ and consider the swapping feature associated with $s$:
\begin{equation*}
    T_i^{swap(s)} := t\big((\mathbf{x}_i', \tilde{\mathbf{x}}_i')_{swap(s)},Y_i'\big).
\end{equation*}
The construction of $T_i$ assures that $\epsilon \odot T_i =T_i^{swap(s)} $ and Lemma \ref{exchangability} implies $T_i \overset{d}= T_i^{swap(s)}$.
The desired result follows by combining the above equations.
\end{proproof2}

The following proposition  is used in our proof of Lemma 4.
\setcounter{proposition}{2}
\begin{proposition}\label{Pro_3}
Suppose that  $||\widehat{\mathbf{G}} - \mathbf{G} ||_2 \leq a_{n_1}$ and
\begin{eqnarray*}
	\frac{1}{C_2} &\leq&  \min \big\{ \lambda_{\min}(\mathbf{G}),  \lambda_{\min}(\mathbf{\widehat{G}}) \big\}\\
&\leq&  \max \big\{ \lambda_{\max}(\mathbf{G}),  \lambda_{\max}(\mathbf{\widehat{G}}) \big\} \leq C_2. \end{eqnarray*}
Then, we have
\begin{equation*}
 \big|e^{-\frac{1}{2}(\mathbf{x},\tilde{\mathbf{x}})\mathbf{G}^{-1} (\mathbf{x},\tilde{\mathbf{x}})^T} - e^{-\frac{1}{2}(\mathbf{x},\tilde{\mathbf{x}})\mathbf{\widehat{G}}^{-1} (\mathbf{x},\tilde{\mathbf{x}})^T}\big| \leq O(a_{n_1})\end{equation*}
 and
 \begin{equation*}
|det(\mathbf{G}) - det(\widehat{\mathbf{G}}) |\leq O(a_{n_1}).\end{equation*}
\end{proposition}

\begin{proproof3}
The proof steps used here are inspired by the proof of Proposition 1 in \cite{Knockoff_RANK}. Based on the matrix norm inequality, we have
\begin{align*}
&\big|(\mathbf{x},\tilde{\mathbf{x}}) \mathbf{G}^{-1} (\mathbf{x},\tilde{\mathbf{x}})^T - (\mathbf{x},\tilde{\mathbf{x}}) \mathbf{\widehat{G}}^{-1} (\mathbf{x},\tilde{\mathbf{x}})^T\big|\\
=& \big|\big|(\mathbf{x},\tilde{\mathbf{x}}) \mathbf{G}^{-1} (\mathbf{x},\tilde{\mathbf{x}})^T - (\mathbf{x},\tilde{\mathbf{x}}) \mathbf{\widehat{G}}^{-1} (\mathbf{x},\tilde{\mathbf{x}})^T\big| \big|_2\\
\leq& \big| \big|(\mathbf{x},\tilde{\mathbf{x}})\big|\big|_2 \cdot \big| \big|  \mathbf{G}^{-1} (\mathbf{x},\tilde{\mathbf{x}})^T - \mathbf{\widehat{G}}^{-1} (\mathbf{x},\tilde{\mathbf{x}})^T\big|\big|_2\\
\leq &\big| \big|(\mathbf{x},\tilde{\mathbf{x}})\big|\big|_2 \cdot \big| \big|  \mathbf{G}^{-1} - \mathbf{\widehat{G}}^{-1} \big|\big|_2 \cdot \big| \big|(\mathbf{x},\tilde{\mathbf{x}})^T\big|\big|_2\\
\leq& \big| \big|(\mathbf{x},\tilde{\mathbf{x}})\big|\big|_2 \cdot \big| \big|(\mathbf{x},\tilde{\mathbf{x}})^T\big|\big|_2\\  &\cdot\big| \big|  \mathbf{G}^{-1}\big| \big|_2 \cdot \big| \big|  \mathbf{\widehat{G}} -\mathbf{G}  \big|\big|_2\cdot \big|\big| \mathbf{\widehat{G}}^{-1} \big|\big|_2\\
\leq & O(a_{n_1}).
\end{align*}
Then,
\begin{align*}
&\big|e^{-\frac{1}{2}(\mathbf{x},\tilde{\mathbf{x}})\mathbf{G}^{-1} (\mathbf{x},\tilde{\mathbf{x}})^T} - e^{-\frac{1}{2}(\mathbf{x},\tilde{\mathbf{x}})\mathbf{\widehat{G}}^{-1} (\mathbf{x},\tilde{\mathbf{x}})^T}\big|\\
  \leq& e^{-\frac{1}{2}(\mathbf{x},\tilde{\mathbf{x}})\mathbf{G}^{-1} (\mathbf{x},\tilde{\mathbf{x}})^T} \cdot \big|1 - e^{O(a_{n_1})}\big|\\
 \leq& O(a_{n_1}).
\end{align*}
This proves the first statement of Proposition \ref{Pro_3}.

Let $\mathbf{G} = (g_{ij})_{1\leq i,j\leq p}$ and $\mathbf{\widehat{G}} = (\hat{g}_{ij})_{1\leq i,j\leq p}$.
Clearly $|g_{ij} - \hat{g}_{ij}|\leq O(a_{n_1})$. Denote $\mathbf{G}_{k}=(g_{ij}^k)_{1\leq i,j\leq k}$ be the $k$-order submatrix of $\mathbf{G}$. Suppose inductively that $|det(\mathbf{G}_{k}) - det(\mathbf{\widehat{G}}_{k})| \leq O(a_{n_1})$ and let $\mathbf{G}^{i1}_{k}$ be the cofactor of $\mathbf{G}_{k+1}$ with respect to $g_{i1}^{k+1}$. Moreover, we have
\begin{align*}
&\big|det(\mathbf{G}_{k+1}) - det(\mathbf{\widehat{G}}_{k+1})\big| \\
=& \big|\sum\limits_{i = 1}^{k+1} \left(g^{i1}_{k} \mathbf{G}^{i1}_{k} - \hat{g}_{i1}^{k+1}\widehat{\mathbf{G}}^{i1}_{k}\right)\big|\\
\leq & \big|\sum\limits_{i = 1}^{k+1} \left((g_{i1}^{k+1} - \hat{g}_{i1}^{k+1}){\mathbf{G}}^{i1}_{k}\right)\big|+ \big|\sum\limits_{i = 1}^{k+1} \left(\hat{g}_{i1}^{k+1}({\mathbf{G}}^{i1}_{k} - \widehat{\mathbf{G}}^{i1}_{k})\right)\big|\\
\leq & O(a_{n_1}).
\end{align*}
Thus, the second desired result follows by the principle of induction.

\end{proproof3}

\begin{prolemma4}
From Proposition 3, we can deduce that
\begin{align*}
&\big| \eta_{\mathbf{\Sigma}}(\mathbf{x},\tilde{\mathbf{x}})- \eta_{\mathbf{\widehat{\Sigma}}}(\mathbf{x},\tilde{\mathbf{x}}) \big| \\
=& \left| \frac{e^{-\frac{1}{2}(\mathbf{x},\tilde{\mathbf{x}}) \mathbf{G}^{-1} (\mathbf{x},\tilde{\mathbf{x}})^T}}{\sqrt{(2\pi)^{2p}\cdot det(\mathbf{G})}}  - \frac{e^{-\frac{1}{2}(\mathbf{x},\tilde{\mathbf{x}}) \mathbf{\widehat{G}}^{-1} (\mathbf{x},\tilde{\mathbf{x}})^T}}{\sqrt{(2\pi)^{2p}\cdot det(\mathbf{\widehat{G}})}} \right|\\
\leq&\frac{\left|\sqrt{det(\mathbf{\widehat{G}})} \big( e^{-\frac{1}{2}(\mathbf{x},\tilde{\mathbf{x}}) \mathbf{\widehat{G}}^{-1} (\mathbf{x},\tilde{\mathbf{x}})^T} -  e^{-\frac{1}{2}(\mathbf{x},\tilde{\mathbf{x}}) \mathbf{G}^{-1} (\mathbf{x},\tilde{\mathbf{x}})^T} \big) \right|}{\sqrt{(2\pi)^{2p}det(\mathbf{\widehat{G}})\cdot det(\mathbf{G})}}\\
& +\frac{\left|( \sqrt{det(\mathbf{\widehat{G}})} - \sqrt{det(\mathbf{G})}) \cdot e^{-\frac{1}{2}(\mathbf{x},\tilde{\mathbf{x}}) \mathbf{G}^{-1} (\mathbf{x},\tilde{\mathbf{x}})^T}  \right|}{\sqrt{(2\pi)^{2p}det(\mathbf{\widehat{G}})\cdot det(\mathbf{G})}}\\
\leq & O(a_{n_1}),
\end{align*}
where the last two inequalities hold with  probability at least $1 - p^{-\frac{1}{b_{n_1}}}$.
%
%

\end{prolemma4}

\begin{proof4}
To derive this claim, it's sufficient to establish that, for any given filter $f$,
\begin{equation*}
\widehat{\text{$k$-FWER}} \leq \alpha + O(p^{-\frac{1}{b_{n_1}}})+ O(a_{n_1}).
\end{equation*}
For any fixed filter $f$, denote $\Gamma$ a set of data matrix satisfying $|\widehat{\mathcal{S}} \cap \mathcal{S}_0| \geq k$ iff $([\mathbf{X}',\mathbf{\tilde{X}}'], \mathbf{y}') \in \Gamma$. In terms of  the conditional independence of knockoff feature, we have
\begin{align*}
k\text{-FWER} &= Prob\{\big( [\mathbf{X}', \mathbf{\tilde{X}}'], \mathbf{y}'\big) \in \Gamma\}\\
 &= \int\limits_{\Gamma} \prod_{i=1}^{n_2} \bigg( \eta(Y_i'|\mathbf{x}_i')\eta_{\mathbf{\Sigma}}(\mathbf{x}_i',\tilde{\mathbf{x}}_i')\bigg) \\
 & \leq \alpha.
\end{align*}
Let $(\mathbf{\tilde{X}}')^{\widehat{\mathbf{\Sigma}}}$ be the Knockoff data matrix constructed via $\widehat{\mathbf{\Sigma}}$. Then, $\widehat{\text{$k$-FWER}}$ can be rewritten as
\begin{align*}
\widehat{\text{$k$-FWER}} &= Prob\{\big( [\mathbf{X}', (\mathbf{\tilde{X}}')^{\widehat{\mathbf{\Sigma}}}], \mathbf{y}'\big) \in \Gamma\}\\
&=\int\limits_{\Gamma} \prod_{i=1}^{n_2} \bigg( \eta(Y_i'|\mathbf{x}_i')\eta_{\widehat{\mathbf{\Sigma}}}(\mathbf{x}_i',\tilde{\mathbf{x}}_i')\bigg).
\end{align*}

Set $\mathcal{A}=\{\eta_{\mathbf{\widehat{\Sigma}}}:\big| \eta_{\mathbf{\Sigma}}(\mathbf{x},\tilde{\mathbf{x}})- \eta_{\mathbf{\widehat{\Sigma}}}(\mathbf{x},\tilde{\mathbf{x}}) \big| \leq O(a_{n_1})\}$. We can deduce that
\begin{small}
\begin{align*}
\widehat{k\text{-FWER}} \bigg| \eta_{\mathbf{\widehat{\Sigma}}} \in \mathcal{A}
 \leq & \int\limits_{\Gamma} \prod_{i=1}^{n_2} \bigg( \eta(Y_i'|\mathbf{x}_i') \cdot \big(\eta_{\mathbf{\Sigma}}(\mathbf{x}_i',\tilde{\mathbf{x}}_i') + O(a_{n_1})\big)\bigg)\\
 \leq & \alpha + O(a_{n_1}).
\end{align*}
\end{small}
Moreover,
\begin{align*}
\widehat{\text{$k$-FWER}} \leq & \big(\widehat{k\text{-FWER}} \big|\eta_{\mathbf{\widehat{\Sigma}}} \in \mathcal{A} \big)\cdot Prob\{\eta_{\mathbf{\widehat{\Sigma}}} \in \mathcal{A}\}\\
& + 1 \cdot Prob\{\eta_{\mathbf{\widehat{\Sigma}}} \not\in \mathcal{A}\}\\
\leq & \alpha + O(a_{n_1}) + O(p^{-\frac{1}{b_{n_1}}}).
\end{align*}
This completes the proof.

\end{proof4}

We omit the proof of Theorem 5 here since its steps are similar with Theorem 4.

\begin{proof6}
denote the set of informative features as $$\mathcal{S}_0 = \{d_1,\dots,d_{|\mathcal{S}_0|}\}$$
and denote
$$\epsilon_{d_j}=Prob\{\xi_{d_j} > \xi\} - 0.5.  $$
For $n_2 \rightarrow \infty$, the central limit theorem implies
\begin{equation}\label{W_informative}
W_{d_j} \overset{d}\longrightarrow \mathcal{N}(\epsilon_{d_j},\sigma_{d_j}^2),
\end{equation}
where $\sigma_{d_j}^2 \leq \frac{0.25}{n_2}$.

 Hence, we have
\begin{align*}
Prob\{W_{d_j} > \frac{\epsilon_{d_j}}{2} \} > & Prob\{\frac{\epsilon_{d_j}}{2} < W_{d_j} < \frac{3\epsilon_{d_j}}{2}\}\\
= & Prob\{ |W_{d_j} - \epsilon_{d_j}| < \frac{\epsilon_{d_j}}{2}\}\\
\geq & 1 - \frac{4\sigma_{d_j}^2}{\epsilon_{d_j}^2}\\
\geq &  1 - \frac{1}{n_2 \cdot \epsilon_{d_j}^2}
\end{align*}
and
$Prob\{W_{d_j} > \frac{\epsilon_{d_j}}{2}\}\rightarrow 1$ as
as $n_2 \rightarrow \infty$.

The power ability of E-Knockoff ($k$-FWER) satisfies that,
\begin{align*}
Power = & \mathbb{E}\left[\frac{|\widehat{\mathcal{S}} \cap \mathcal{S}_0|}{|\mathcal{S}_0|} \right]\\
\geq & \frac{1}{|\mathcal{S}_0|} \sum\limits_{j = 1}^{|\mathcal{S}_0|}Prob\{\mathcal{P}_{d_j} \leq \alpha_1 \}\\
\geq  &  \frac{1}{|\mathcal{S}_0|} \sum\limits_{j = 1}^{|\mathcal{S}_0|}Prob\{W_{d_j} \geq \text{icdf}(1 - \frac{k\alpha}{2p})\cdot \sqrt{\frac{0.25}{n_2}} \}\\
\geq  &  \frac{1}{|\mathcal{S}_0|} \sum\limits_{j = 1}^{|\mathcal{S}_0|}Prob\{W_{d_j} \geq \frac{\epsilon_{d_j}}{2} \}\\
\rightarrow & 1~~ \mbox{as}~~ n_2 \rightarrow \infty,
\end{align*}
 where icdf is the inverse cumulative distribution function of $\mathcal{N}(0,1)$.

 The power analysis for E-Knockoff (FDP) also can be obatined by the similar analysis as above. For simplicity, we omit it here.

For E-Knockoff (FDR), the central limit theorem implies that,
\begin{equation}\label{W_irrelevant}
W_j \overset{d}\rightarrow \mathcal{N}(0, 0.25/n_2), \forall j \in \mathcal{S}_1.
\end{equation}
Let $\epsilon = \text{min}\{\epsilon_{d_j}, j = 1,\dots, |\mathcal{S}_0|\}$.  According to \eqref{W_informative} and \eqref{W_irrelevant}, we get
\begin{align*}
&Prob\{\min_jW_j \leq -\epsilon/2\} \\
= & Prob\{W_1 \leq -\epsilon/2 \lor \dots \lor W_p \leq -\epsilon/2\}\\
\leq & \sum\limits_{j = 1}^p Prob\{W_j \leq -\epsilon/2\}\end{align*}
and
$Prob\{\min_jW_j \leq -\epsilon/2\}
\rightarrow  0$ as $n_2 \rightarrow \infty$.

By the construction of threshold value $\tau$ (see also Lemma 1 in main paper), we can deduce that $\tau \leq \max\{0, -\min_jW_j\}$.
Thus, the power of E-knockoff (FDR) satisfies
\begin{small}
\begin{align*}
Power \geq & \mathbb{E}\left[\frac{|\widehat{\mathcal{S}} \cap \mathcal{S}_0|}{|\mathcal{S}_0|} \bigg| \min_jW_j > -\epsilon/2\right] \cdot Prob\{\min_jW_j > -\epsilon/2 \}\\
\geq  &  \frac{1}{|\mathcal{S}_0|} \sum\limits_{j = 1}^{|\mathcal{S}_0|}Prob\{W_{d_j} > \epsilon/2 \} \cdot Prob\{\min_jW_j > -\epsilon/2 \}\\
\rightarrow & 1
\end{align*}
\end{small}
as $n_2 \rightarrow \infty.$

\end{proof6}

\begin{table}[!htbp]
\renewcommand\arraystretch{1.3}
\centering
\caption{Simulation results for robust analysis}\label{Simulation_robust}
\resizebox{8cm}{!}{
\begin{tabular}{c|cc|cc|cc}
\Xhline{1.1pt}
\multirow{2}*{\tabincell{c}{$n_1$}}&\multicolumn{2}{c|}{E-Knockoff($k$-FWER)}&\multicolumn{2}{c|}{E-Knockoff(FDP)}&\multicolumn{2}{c}{E-Knockoff(FDR)}\\
&FD$_{\max}$&Power&FDP$_{\max}$&Power&FDR&Power\\
\Xhline{1.1pt}
100 &1&0.18&0.50&0.19&0.21&0.30\\
200 &3&0.72&0.19&1.00&0.18&0.80\\
300 &2&0.98&0.15&1.00&0.14&1.00\\
400 &2&1.00&0.19&1.00&0.21&1.00\\
500 &1&1.00&0.14&1.00&0.18&1.00\\
600 &2&1.00&0.14&1.00&0.19&1.00\\
700 &2&1.00&0.09&1.00&0.16&1.00\\
800 &1&1.00&0.09&1.00&0.20&1.00\\
900 &2&1.00&0.09&1.00&0.15&1.00\\
1000&1&1.00&0.14&1.00&0.16&1.00\\
\Xhline{1.1pt}
\end{tabular}}
\end{table}

\begin{table}[!htbp]
\renewcommand\arraystretch{1.3}
\centering
\caption{Power analysis for E-Knockoff ($k$-FWER)}\label{Simulation_result_k-FWER}
\resizebox{7.5cm}{!}{
\begin{tabular}{c|cc|cc|cc}
\Xhline{1.1pt}
\multirow{2}*{\tabincell{c}{$n_2$}}&\multicolumn{2}{c|}{DNN}&\multicolumn{2}{c|}{Lasso}&\multicolumn{2}{c}{KRR}\\
&FD$_{\max}$&Power&FD$_{\max}$&Power&FD$_{\max}$&Power\\
\Xhline{1.1pt}
200 &2&0.10&1&0.06&1&0.01\\
400 &1&0.36&1&0.19&2&0.02\\
600 &1&0.63&2&0.34&2&0.04\\
800 &1&0.78&1&0.50&1&0.06\\
1000&1&0.90&1&0.64&1&0.10\\
1200&1&0.96&2&0.74&2&0.13\\
1400&1&0.99&2&0.81&2&0.17\\
1600&1&1.00&2&0.87&2&0.23\\
1800&1&1.00&1&0.91&2&0.29\\
2000&1&1.00&1&0.94&3&0.26\\
\Xhline{1.1pt}
\end{tabular}}
\end{table}

\section{Supplementary Material D}

\subsection{D.1 Implementation Details}
For kernel ridge regressor (KRR), we use sigmoid function as non-linear kernel (gamma $1/p$, bias coefficient $1.0$) \cite{sklearn_api}. In the Lasso case, the max number of iterations and the optimization tolerance are set at $1000$ and $1e-4$, respectively. The regularization coefficient is selected from 100  regularization coefficients by five-fold cross validation \cite{sklearn_api}. For DNN, we use two hidden layers with $p$ nodes \cite{LuDeepPINK}. Both ReLU activation \cite{relu} and $l_1$ regularization (regularization coefficient 0.01) are employed in this network. We use Adam optimizer \cite{adamopt} (learning rate 0.001, beta1 0.9, beta2 0.99, epsilon $1e-07$) with respect to mean square error loss to train the network \cite{tensorflow2015-whitepaper}. The batch size and the number of epochs are set at 100 and 500, respectively.
For the paired-input DNN described in \cite{LuDeepPINK}, we apply batch size 100 and number of epochs 500. Other settings are followed from \cite{LuDeepPINK}.

\begin{table}[!htbp]
\renewcommand\arraystretch{1.3}
\centering
\caption{Power analysis for E-Knockoff (FDP)}\label{Simulation_result_FDP}
\resizebox{8cm}{!}{
\begin{tabular}{c|cc|cc|cc}
\Xhline{1.1pt}
\multirow{2}*{\tabincell{c}{$n_2$}}&\multicolumn{2}{c|}{DNN}&\multicolumn{2}{c|}{Lasso}&\multicolumn{2}{c}{KRR}\\
&FDP$_{\max}$&Power&FDP$_{\max}$&Power&FDP$_{\max}$&Power\\
\Xhline{1.1pt}
200 &1.00&0.12&1.00&0.07&1.00&0.00\\
400 &0.16&0.48&0.25&0.29&1.00&0.01\\
600 &0.15&0.78&0.21&0.49&1.00&0.03\\
800 &0.10&0.91&0.13&0.67&1.00&0.06\\
1000&0.13&0.96&0.12&0.81&1.00&0.13\\
1200&0.15&0.99&0.09&0.89&0.33&0.19\\
1400&0.09&1.00&0.11&0.94&0.20&0.32\\
1600&0.14&1.00&0.10&0.96&0.22&0.42\\
1800&0.12&1.00&0.09&0.98&0.18&0.49\\
2000&0.12&1.00&0.12&0.99&0.17&0.56\\
\Xhline{1.1pt}
\end{tabular}}
\end{table}

\begin{table}[!htbp]
\renewcommand\arraystretch{1.3}
\centering
\caption{Power analysis for E-Knockoff (FDR)}\label{Simulation_result_FDR}
\resizebox{7.5cm}{!}{
\begin{tabular}{c|cc|cc|cc}
\Xhline{1.1pt}
\multirow{2}*{\tabincell{c}{$n_2$}}&\multicolumn{2}{c|}{DNN}&\multicolumn{2}{c|}{Lasso}&\multicolumn{2}{c}{KRR}\\
&\ FDR\ &Power&\ FDR\ &Power&\ FDR\ &Power\\
\Xhline{1.1pt}
200 &0.21&0.23&0.24&0.32&0.58&0.02\\
400 &0.18&0.71&0.17&0.60&0.45&0.03\\
600 &0.17&0.91&0.14&0.78&0.19&0.09\\
800 &0.17&0.96&0.17&0.91&0.22&0.14\\
1000&0.16&0.99&0.15&0.94&0.24&0.30\\
1200&0.12&1.00&0.15&0.97&0.25&0.37\\
1400&0.21&1.00&0.18&0.99&0.23&0.47\\
1600&0.20&1.00&0.17&0.99&0.21&0.55\\
1800&0.17&1.00&0.16&1.00&0.23&0.65\\
2000&0.18&1.00&0.17&1.00&0.20&0.70\\
\Xhline{1.1pt}
\end{tabular}}
\end{table}

\begin{table}[htbp]
\renewcommand\arraystretch{1.3}
\centering
\caption{The sample size and dimensions of each drug}\label{real_dataset_detail}
\resizebox{7.5cm}{!}{
\begin{tabular}{c|c|cc}
\Xhline{1.1pt}
Class& drugs &observations ($n$)& mutations ($p$)\\
\Xhline{1.1pt}
\multirow{7}{*}{PI}&ATV&1218&223\\
&FPV&1809&224\\
&NFV&1907&228\\
&IDV&1860&227\\
&LPV&1652&236\\
&SQV&1861&223\\
&TPV&908&249\\
\Xhline{1.1pt}
\multirow{6}{*}{NRTI}
&ABC&1597&379\\
&AZT&1683&378\\
&D4T&1693&379\\
&DDI&1693&378\\
&TDF&1354&378\\
&3TC&1662&373\\
\Xhline{1.1pt}
\multirow{2}{*}{NNRTI}&EFV&1742&371\\
&NVP&1740&370\\
\Xhline{1.1pt}
\end{tabular}}
\end{table}

\setcounter{figure}{3}
\begin{figure*}[!ht]
\centering
\includegraphics[scale=0.37]{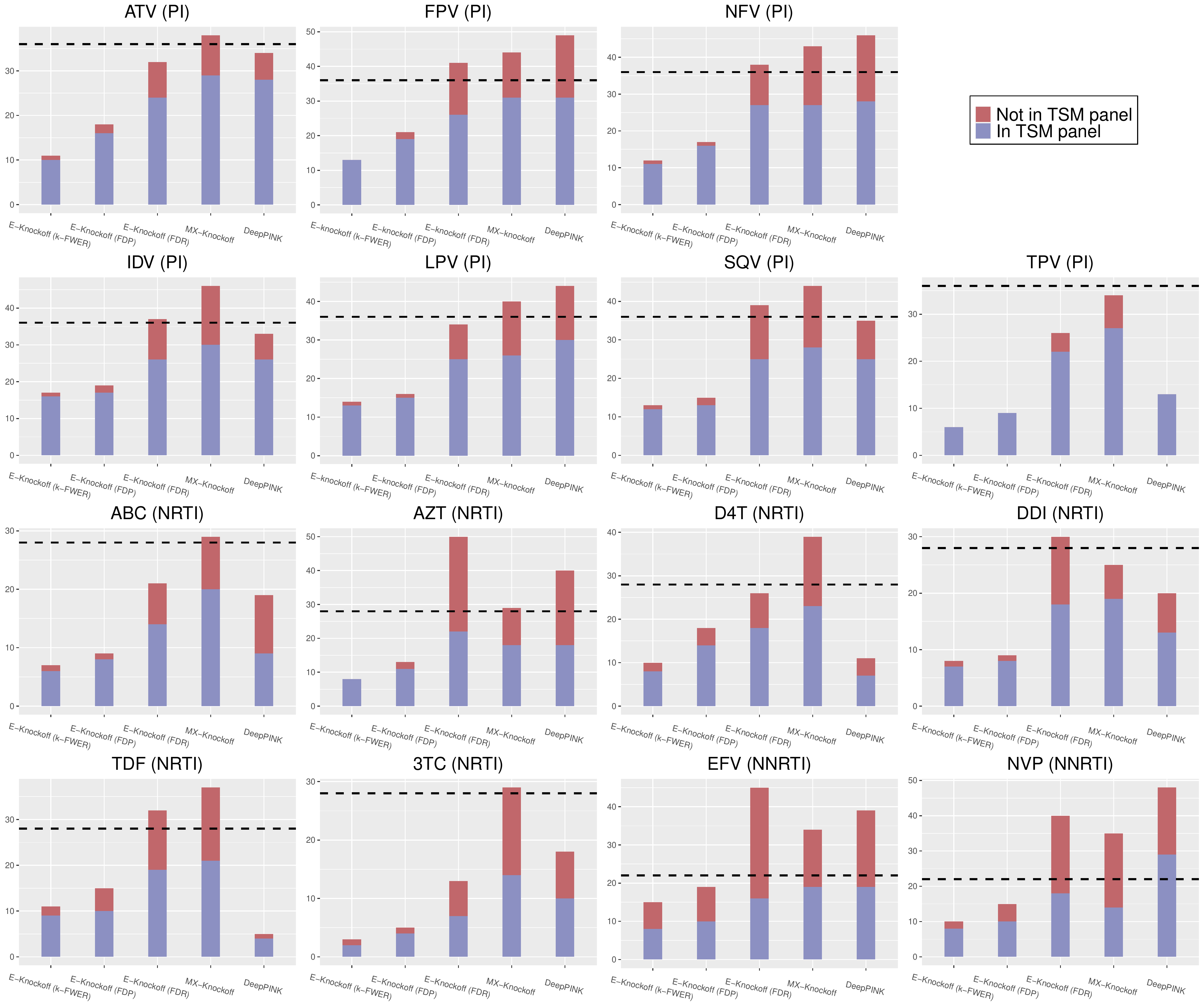}
\caption{Results on the HIV-1 drug resistance dataset. For each drug class, we plot the number of protease positions (for PI) or reverse transcriptase (RT) positions (for NRTI or NNRTI) selected  by different knockoff filters. The color indicates whether or not the selected position appears in the treatment selected mutation (TSM) panel, and the horizontal line shows the total number of positions on the TSM panel. }
\label{HIV_results}
\end{figure*}

\subsection{D.2 Simulations for Robustness }
To valid the  theoretical findings on robustness, an experiment is provided to illustrate the influence of $n_1$ on selection results when the argument follows some unknown Gaussian Graphical model. We set $p=2000$, $n_2=1000$ and select $n_1$ from $\{100,200,300, 400,500,600,700,800,900,1000\}$.
After repeating each experiment 30 times, the max number of false discoveries ($\text{FD}_{\max}$), $\text{FDP}_{\max}$, FDR and Power of these simulations are reported in Table \ref{Simulation_robust}.
It can be observed that $\text{FD}_{\max}$, $\text{FDP}_{\max}$ and FDR can be better controlled with the increase of $n_1$, which is consistent with Theorems 4 and 5.

\subsection{D.3 Full Experimental Results for Power analysis}
Full experimental results for power analysis is provided in Tables \ref{Simulation_result_k-FWER}-\ref{Simulation_result_FDR}.  When keeping the $\text{FD}_{\max}, \text{FDP}_{\max}$ or FDR under control,  we can see that, in most of the settings, a powerful selection result can be made when more samples are added to the second part of dataset. This supports our conclusion in Theorem 6.
Indeed, some  $\text{FD}_{\max}$ or $\text{FDP}_{\max}$  may exceed the fixed level since $k$-FWER control and FDP control allow errors with a certain probability.

\subsection{D.4 Experimental Analysis for HIV Dataset}
The HIV-1 drug resistance dataset \cite{hivdataset} contains data for eight protease inhibitor (PI) drugs, six nucleoside reverse transcriptase inhibitor (NRTI) drugs and four nonnucleoside reverse transcriptase inhibitor (NNRTI) drugs. The response $Y$ is the log-transformed drug resistance level while the feature $X_j$ are the markers for the presence or absence of the $j$th mutation \cite{Knockoff_group_linear}.
Inspired by \cite{Knockoff_group_linear}, the drug with high proportion of missing drug resistance measurement are removed (each with over 50\% missing data). We only keep those mutations which appear $>10$ times in the sample set,  and report the characteristics of  resulting dataset  in Table \ref{real_dataset_detail}.

We then apply the E-Knockoff inference associated with Lasso ($2n_1 =  n_2 = \frac{2n}{3}$), including E-Knockoff ($k$-FWER) ($k=2, \alpha = 0.1$), E-Knockoff (FDP) ($q = 0.2, \alpha = 0.2$), and E-Knockoff (FDR) ($q = 0.2$), to these datasets. For comparison, we also apply the Lasso-based MX-Knockoff ($q = 0.2$) and DeepPINK ($q = 0.2$) to the same data.
Figure \ref{HIV_results} summaries the selected mutations against the treatment selected mutations (TSM).

Experimental resuts show that E-Knockoff can control the false discovery efficiently for seven PI drugs. Also, the proposed methods would produce many ``false discoveries'' in NRTI and NNRTI drugs. This may due to the fact that TSM panel may not contain all informative mutations. Similar experimental results are also reported in \cite{Knockoff_group_linear,LuDeepPINK}

\end{document}